\documentclass[fleqn,10pt,twocolumn]{wlscirep}
\usepackage[utf8]{inputenc}
\usepackage[T1]{fontenc}

\usepackage{multicol}
\usepackage{blindtext}

\usepackage{nccmath}
\usepackage{latexsym}
\usepackage{amssymb}
\usepackage{amsmath}
\usepackage{amsfonts}

\usepackage{textcase}

\usepackage{siunitx}
\usepackage{graphics}
\usepackage{float}
\graphicspath{{Figure/}}

\usepackage{kantlipsum}

\usepackage{lineno}

\usepackage[version=3]{mhchem}

\usepackage{booktabs}
\usepackage{multirow}

\usepackage{textcomp}
\usepackage{gensymb}

\usepackage{xparse}

\begin{document}
\title{High-\emph{Q} longwave infrared microresonators based on a non-epitaxial germanium platform}
\author[1,2,*]{Dingding Ren}
\author[1]{Chao Dong}
\author[1]{David Burghoff}

\affil[1]{Department of Electrical Engineering, University of Notre Dame, 46556 IN, USA}
\affil[2]{Department of Electronic Systems, Norwegian University of Science and Technology (NTNU), 7491 Trondheim, Norway}

\affil[*]{e-mail: dren@nd.edu}

%\affil{\mbox{}}

\makeatletter
\renewcommand{\@maketitle}{%
{%
\thispagestyle{empty}%
\vskip-36pt%
{\raggedright\sffamily\bfseries\fontsize{20}{25}\selectfont \@title\par}%
\vskip10pt
{\raggedright\sffamily\fontsize{12}{16}\selectfont  \@author\par}
\vskip25pt%
}%
}%
\makeatother

%\author[1,$\dag$]{Dingding Ren}
%\author[2,$\dag$]{Chao Dong}
%\author[1,2]{David Burghoff}
%\author[2,*]{Derek Author}

\maketitle
\newif\ifpreprint
\preprinttrue % uncomment for preprint

% Use commas for non-preprint, periods for preprint
\ifpreprint
\newcommand\figsub[1]{\textbf{#1}.}
\else
\newcommand\figsub[1]{\textbf{#1},}
\fi

% Do not call out Supplementary Figures for preprint
\ifpreprint
\newcommand\suppfig[1]{}
\else
\newcommand\suppfig[1]{ (Supplementary Fig. #1)}
\fi

% \introcite will include references in the abstract if not the preprint
\ifpreprint
\newcommand\introcite[1]{}
\else
\DeclareDocumentCommand{\introcite}{m o o o o o o o o}
{%
\IfNoValueTF{#9}{%
    \IfNoValueTF{#8}{%
        \IfNoValueTF{#7}{%
            \IfNoValueTF{#6}{%
                \IfNoValueTF{#5}{%
                    \IfNoValueTF{#4}{%
                        \IfNoValueTF{#3}{%
                            \IfNoValueTF{#2}{%
                                \cite{#1}}%
                                {\cite{#1,#2}}}%
                            {\cite{#1,#2,#3}}}%
                        {\cite{#1,#2,#3,#4}}}%
                    {\cite{#1,#2,#3,#4,#5}}}%
                {\cite{#1,#2,#3,#4,#5,#6}}}%
            {\cite{#1,#2,#3,#4,#5,#6,#7}}}%
        {\cite{#1,#2,#3,#4,#5,#6,#7,#8}}}%
    {\cite{#1,#2,#3,#4,#5,#6,#7,#8,#9}}%
}%
\fi

\newcommand\waferbonding{%
A piece of SD-2 glass of $1.5 \times 1.5\,\si{cm^{2}}$ in size and 1 mm in thickness was firstly cleaned by acetone and isopropanol. This was followed by a hot bath in \ce{H_{2}O_{2}}:\ce{NH_{4}OH}:\ce{H_{2}O} (volume ratio of 1:1:5) at 90\celsius, for 15 minutes. This step ensures that the surface of the glass has high surface energy, favorable for hydrophilic bonding\cite{grudinin2016properties}. A piece of highly resistive Ge of $1 \times 1\, \si{\cm^2}$ in size and 500 \micro m in thickness (purchased from Umicore) was dipped into the same \ce{H_{2}O_{2}}:\ce{NH_{4}OH}:\ce{H_{2}O} solution at room temperature for 5 seconds. Any  surface treatment of Ge at a higher temperature or for a longer time will lead to surface degradation. After the surface treatment, the glass and Ge piece were sandwiched into graphite plates inside a homemade bonder. 3000 N force was applied, and the bonder was heated to 300 \celsius\,for 60 min\suppfig{S1}.}

\newcommand\nanofab{%
After the mechanical lapping and polishing process, electron beam lithography (EBL) was used to define the features of the Ge, using a Vistec EBPG5200 EBL System at 100 KV. HMDS adhesion promoter and Espacer-300Z discharger were applied before and after the spin-coating of the e-beam resist ma-N 2403, this is essential. An SF$_6$ chemistry in a deep RIE system was used to etch through the Ge layer to define the feature of the microresonator. To ensure all the facets of the resonator were finished with a polished smooth surface, the etched microresonator was polished after the nanofabrication. The sample was then iteratively etched using HF acid etching alternating with a sidewall passivation process; this allows us to etch back the glass layer from the edge of the resonator. This was performed three times with 49\% HF for 1 min in each etching step. A flood exposure with positive S1813 photoresist has been used to passivate the vertical side wall, using the microresonator itself as the passivation mask. This iteration of etching and passivation prevented severe in-plane etching that could easily wash away the microresonator and it maintained effective etching speed in the vertical direction. We fabricated the  curved Ge waveguides using the same fabrication process in order to couple the light from free space into the microresonator. The Ge waveguide has a larger in-plane end facet to enhance the collection of the light from the free space. Both the coupling sections of the Ge microresonator and the waveguide were diced to the edge of the glass chips. A final HF etching process has been used to further etch back the glass, suspending both the Ge microresonator and the Ge waveguide from the glass chips.}

\newcommand\dfbtuning{%
A DFB QCL with peak power of 250 mW and wavelength tuning range from 1285.5 to 1290.5 $\si{\cm^{-1}}$ has been used to perform the wavelength-dependent transmission measurements. A linear parameter space of current and temperature was established in order to wavelength-tune at a constant power of 10 mW. An Arroyo 6310-QCL controller was used to control the temperature of the TEC and current of the QCL laser with a minimum step size of 0.01 $\si{\celsius}$ and 0.02 $\si{\mA}$.}

\textbf{The longwave infrared (LWIR) region of the spectrum spans 8 to 14 \micro m and enables high-performance sensing and imaging for detection, ranging, and monitoring\introcite{vollmer2017infrared,kaplan2007practical}. Chip-scale integrated LWIR photonics has enormous potential for real-time environmental monitoring, explosive detection, and biomedicine. However, realizing advanced technologies such as precision sensors\introcite{armani2007label} and broadband frequency combs\introcite{kippenberg2018dissipative} requires ultra low-loss components, which have so far remained elusive in this regime.\introcite{wilson2019intense,lesko2021six,cao2020towards,della2021mid} 
\ifpreprint
We demonstrate
\else
Here, we demonstrate
\fi that non-epitaxial germanium is an enabling technology for longwave infrared integrated photonics. We use it to demonstrate the first high quality (\emph{Q}) factor whispering gallery mode microresonators in the LWIR, which we couple to integrated low-loss waveguides. At 8 \micro m, we measure losses of 0.5 dB/cm and intrinsic \emph{Q} factors of $2.5 \times 10^{5}$, nearly two orders of magnitude higher than prior LWIR resonators.\introcite{ramirez2019broadband,lee2020high,kozak2021germanium} Our work portends the development of integrated sensing and nonlinear photonics in the LWIR regime.}

% If you want to add this stuff, put it somewhere else
% In addition, the rovibrational transition of molecules in LWIR can be orders of magnitude stronger than near-infrared wavelengths with fingerprints of compounds such as \ce{C_{2}H_{4}}\cite{brown2003absorption}, TNT\cite{webber2005optical}, and glucose\cite{kasahara2018noninvasive,jernelv2019review}.
% However, the spectroscopy techniques at LWIR rely on high-power laser source, like \ce{CO_{2}} laser or quantum cascade laser (QCL), with a significant amount of input power, and there are 

Dielectric optical microresonators are an essential tool for nonlinear optics. By storing high energy in a small volume, they significantly enhance nonlinear light-matter interactions. A number of interesting effects can be demonstrated in them: frequency comb generation\cite{delhayeOpticalFrequencyComb2007,gaeta2019photonic}, ultra-sensitive sensing\cite{armani2007label,guggenheim2017ultrasensitive}, optomechanics\cite{anetsberger2008ultralow,wiederhecker2009controlling}, and non-Hermitian physics\cite{peng2014parity,chang2014parity,cao2015dielectric}, to name a few.
%chip-scale  dispersion engineering\cite{yang2016broadband,kim2017dispersion},
In the near-infrared, platforms such as \ce{SiO_{2}} microtoroids \cite{armaniUltrahighToroidMicrocavity2003} and \ce{Si_{3}N_{4}} ring resonators\cite{pfeiffer2018ultra} can have \emph{Q} factors of billions, while at longer-wavelength midwave infrared,  \emph{Q} values can be sustained at the level of several million \cite{luke2015broadband}. Still, there is significant interest in pushing to even longer wavelengths. The rovibrational transition of molecules in the LWIR can be orders of magnitude stronger than shorter wavelengths \cite{brown2003absorption,webber2005optical,kasahara2018noninvasive,jernelv2019review}, and of course most blackbody radiation occurs in this range, which is why forward-looking infrared imaging has become so prominent. While high-power sources like CO$_2$ lasers and quantum cascade lasers\cite{hugiMidinfraredFrequencyComb2012,mengBroadlyTunableSinglemode2015,luHighPowerFrequency2015,hillbrandCoherentInjectionLocking2019,mengMidinfraredFrequencyComb2020} are readily available, nonlinear integrated photonics remains challenging. The scaling with the wavelength becomes unfavorable and almost all materials become lossy. Chalcogenides become moderately lossy beyond 7 \micro m \cite{fengFewmodedUltralargeMode2020}, and
% Driven by a vigorous application perspective at a longer wavelength for ultrasensitive sensing\cite{vollmer2008whispering} and QCL stabilization\cite{siciliani2016microcavity} and frequency comb generation\cite{savchenkov2015generation}, crystalline microresonators with \emph{Q} factors of tens of millions, such as Si\cite{miller2017low} and fluorides\cite{grudinin2016properties,lecaplain2016mid}, have attracted great attention lately.
silicon has significant absorption loss (over 1dB/cm) beyond 6 \micro m due to multiphonon absorption\cite{millerLowlossSiliconPlatform2017,bendow1977multiphonon}
% , and there is no mature chemistry available for nanofabrication of integrated fluoride-based optoelectronics.
So far, the lack of high-\emph{Q} microresonators and correspondingly low-loss waveguides suitable for LWIR remains a bottleneck for LWIR integrated photonics. Despite substantial efforts\cite{ramirez2019broadband,leeHighQDiamondMicroresonators2020,kozakGermaniumonsiliconWaveguidesLongwave2021}, the \emph{Q}s have been so low that none have even scanned a laser across a resonance and observed clear, narrow features, something routine at shorter wavelengths.

Germanium is a legendary material platform in which the first transistors were realized, and it is fairly compatible with mainstream silicon technology\cite{carrollDirectGapGainOptical2012}. With the highest refractive index ($n_0=4$), a large nonlinear coefficient ($n_2$=20 \si{nm^2/W}), and a wide transparency window (from 2 to 14 \micro m), Ge shows excellent potential as a low-loss material platform at LWIR\cite{zhangNonlinearGroupIV2014,hon2011third,soref2010mid}. To fabricate Ge-based on-chip photonic devices, heterogeneous integration is essential, as microstructures are needed to confine light propagation into a small volume. Recently, epitaxially-grown Ge-on-GaAs and Ge-on-Si has been used to make integrated waveguides and ring resonators \cite{chang2012low,brun2014low,nedeljkovic2017germanium,liaoLowLossGeonGaAsPlatform2017,gallacher2018low,montesinos2020ge,kozakGermaniumonsiliconWaveguidesLongwave2021}. The optical mode was primarily confined in the Ge part, achieving waveguide losses of 4 dB/cm at LWIR. However, these losses are an order of magnitude higher (or even more) than the intrinsic losses of high-quality Ge grown by the traditional Czochralski crystal pulling method under thermodynamic equilibrium conditions. In fact, the losses of pure germanium are so low (<0.15 \si{cm^{-1}}) that high-quality measurements of the losses were only performed recently \cite{leeLongwaveInfraredAbsorption2020}. Although epitaxial Ge can be atomically smooth, whenever GaAs or Si are used as the substrate for heteroepitaxy, inevitable degradation of the LWIR transparency results. For Ge grown on GaAs, the volatile arsenic background and the Ga interdiffusion causes unwanted impurity incorporation, lowering the optical transparency at LWIR\cite{bai2012photoluminescence}. Although Ge grown on Si will not be impacted by the unwanted background impurity incorporation as in GaAs, there is a significant lattice mismatch between Si and Ge. This results in a high density of threading dislocations and significantly lower material quality \cite{eaglesham1991low}. The thick waveguides needed for the LWIR excacerbate this issue, and the high index of silicon allows light to penetrate into silicon.

\begin{figure*}[ht!]
\centering\includegraphics[width=0.8\textwidth]{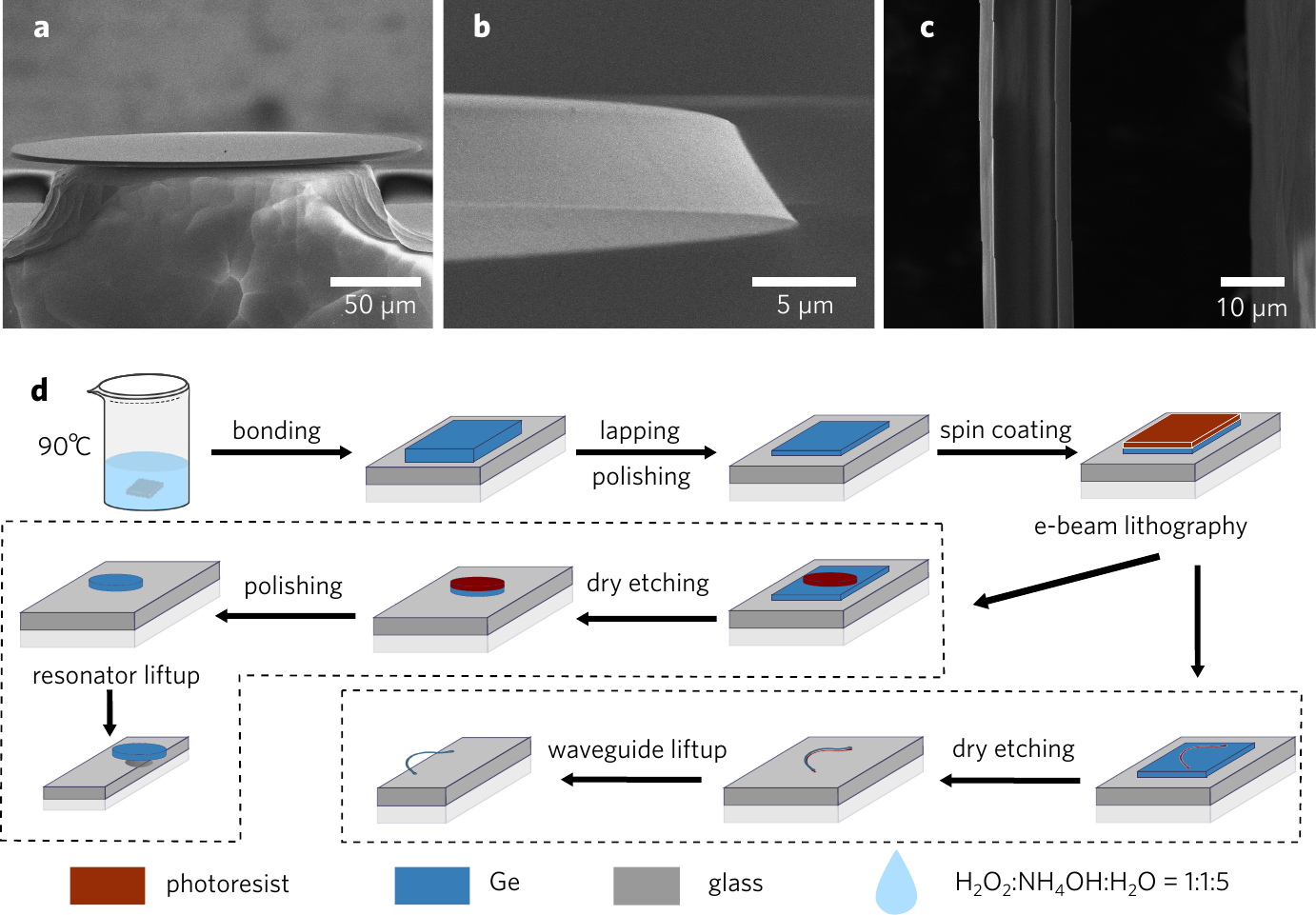}
\caption{\label{fig:fab}\textbf{Images and fabrication of the non-epitaxial Ge microresonators and suspended waveguides.} \figsub{a} SEM image of a non-epitaxial Ge microresonator. The 450 \micro m disk is supported by a glass pillar. \figsub{b} Magnified view of the microresonator edge where the WGM mode resides. \figsub{c} Top view of the suspended Ge waveguide. The waveguide is on the left and the glass substrate is on the right. \figsub{d} Overview of the fabrication procedure, which relies on a combination of wafer bonding and polishing.
% Hydrophilic surface treatment of SD-2 glasses. \textbf{be}, Heterointegration of Ge and glass by wafer bonding. \textbf{c}, Mechanical lapping, and polishing process of bonded Ge to the thickness of 5 \micro m. \textbf{d}, Spin coating of ma-N2403 e-beam resist. \textbf{e}, E-beam lithography to define the feature of WGM. \textbf{f}, Deep reactive ion etching process to form the Ge WGM on glass. \textbf{g}, Final polishing process to smooth the WGM sidewall. \textbf{h}, The detachment of the edge of the Ge WGM from the bonded glass surface using HF acid etching. \textbf{i}, E-beam lithography to define the feature of Ge waveguide. \textbf{j}, Deep reactive ion etching process to form the Ge waveguide on glass. \textbf{k}, suspending the waveguide using HF acid.
}
\end{figure*}

To truly see the benefits of Ge in the LWIR, it is necessary to use the highest-quality, lowest-loss material. Here, we utilize mechanical fabrication processes of wafer bonding, lapping, and polishing to fabricate extremely high-Q whispering gallery mode (WGM) resonators and low-loss waveguides. The structures are initially bonded to glass substrates (essentially creating thick Ge-on-insulator), but this glass is mostly removed, resulting in structures that are are ultimately Ge-on-nothing. These microresonators have ultra-smooth surfaces and are easily separated from their substrates using conventional etchants. To evaluate the optical properties of the resonator, we performed transmission measurements by focusing the emission of an 8 \micro m distributed feedback quantum cascade laser (QCL) into a semi-suspended Ge waveguide, which was coupled into the microresonator. By sweeping the wavelength of the QCL, a loaded \emph{$Q_{l}$} of $2.2 \times 10^{5}$ and an intrinsic \emph{$Q_{i}$} of $2.5 \times 10^{5}$ were measured. This was similar for both the transverse electric (TE) and transverse magnetic (TM) polarizations.

Figure \ref{fig:fab}a-c shows scanning electron microscope (SEM) images of the fabricated devices. A microresonator with diameter of 450 \micro m can be seen in Fig. \ref{fig:fab}a. The resonator has a uniform thickness across the whole resonator, while the outermost 100 \micro m is suspended away from the dune-shaped glass supporting structure in the center. This eliminates the losses due to the glass substrate at LWIR. To look at the active part of the WGM microresonator, we zoomed in to the edge of the WGM microresonator. Fig. \ref{fig:fab}b shows a zoomed image of the edge. It is angled inward by about 30-degrees, a consequence of etching isotropy and edge rounding from the polishing process. By changing the plasma etching anisotropy strength and polishing pressure, the geometry could be further tuned, with potential for dispersion engineering\cite{yangBroadbandDispersionengineeredMicroresonator2016}. To efficiently couple LWIR light into the WGM microresonator, we fabricated a partially-suspended Ge waveguide on a glass chip with dimensions of 4 $\si{\times}$ 2 $\si{\mm^{2}}$, shown in Fig. \ref{fig:fab}c. The waveguide is around 5 \micro m thick, similar to that of the microresonator, and is 10 \micro m wide. The curved part of the suspended waveguide is shown in the figure---its uniform geometry and smooth surface make it ideal for coupling measurements.

Figure \ref{fig:fab}d shows an overview of the fabrication process of the microresonators and of the partially-suspended waveguides.
\ifpreprint Briefly, it relies on a combination of etching and post-polishing. \waferbonding

\nanofab
\else The details are summarized in the Methods, but briefly it relies on a combination of etching and post-polishing. 
\fi
The bottom side of the WGM microresonator has only gone through the wafer bonding process and has the same surface roughness as the original polished Ge wafer, and we did not find a significant difference in the surface roughness between the side-facet and the bottom facet. We performed atomic force microscopy in tapping mode in the center of the WGM microresonator to evaluate the surface roughness of polished Ge surface. Only sparsely distributed spot-like extrusions with height of tens of nanometers could be found over the 13.5 $\si{\times}$ 13.5 $\si{\mm^{2}}$ area\suppfig{S2}. Considering that the surface roughness of polished WGM microresonator is about 2 orders of magnitude smaller than the wavelength in LWIR, so we can safely ensure that there is no significant surface roughness-induced scattering loss for our WGM microresonator at the LWIR.

\begin{figure}[ht!]
    \includegraphics[width=\linewidth]{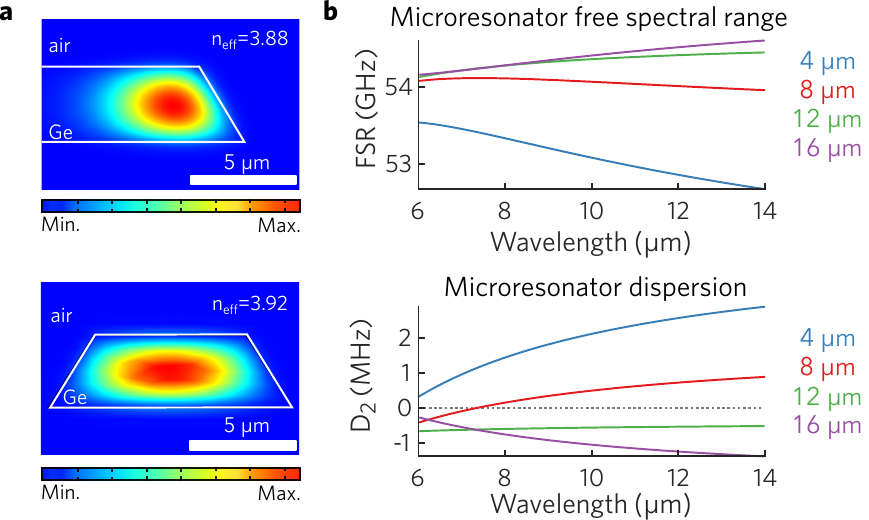}
    \caption{\label{fig:sim}\textbf{Mode simulation and microresonator dispersion.} \figsub{a} FEM simulation of the guided mode in the Ge WGM microresonator and Ge waveguide. \figsub{b} Calculated free spectral range and dispersion versus waveguide thickness. Low anomalous dispersion is only possible with very thick waveguides.}
\end{figure}
To simulate the light propagation in the microresonator and the waveguides, we performed finite element simulations based on the geometry and dimensions shown in Figure \ref{fig:sim}. The effective refractive indices of 3.88 and 3.92 were calculated for the fundamental (quasi-) TE mode in the WGM microresonator and Ge waveguide, respectively, with each mode profile shown in Fig. \ref{fig:sim}a. To explore the tunability of nonlinearity in Ge WGM microresonators, we calculated the thickness-dependent free-spectral range (\emph{FSR}) and \emph{$\Delta$FSR} per mode ($D_{2}$). As is shown in Fig. \ref{fig:sim}b, there is a significant increase of \emph{FSR} with increasing thickness of the WGM microresonator (over the entire spectrum LWIR range). Similarly, with increasing thickness $D_{2}$ the dispersion goes from positive to anomalous dispersion. This tuning range is several orders of magnitude higher than that of the \ce{SiO_{2}} WGM microresonators, a result of the significant refractive index contrast between Ge and air\cite{yangBroadbandDispersionengineeredMicroresonator2016}. Note that achieving low and flat anomolous dispersion with these resonators requires Ge thicknesses greater than 10 \micro m, which would be very difficult using epitaxially-grown samples on non-native substrates.

\begin{figure*}[ht!]
    \centering\includegraphics{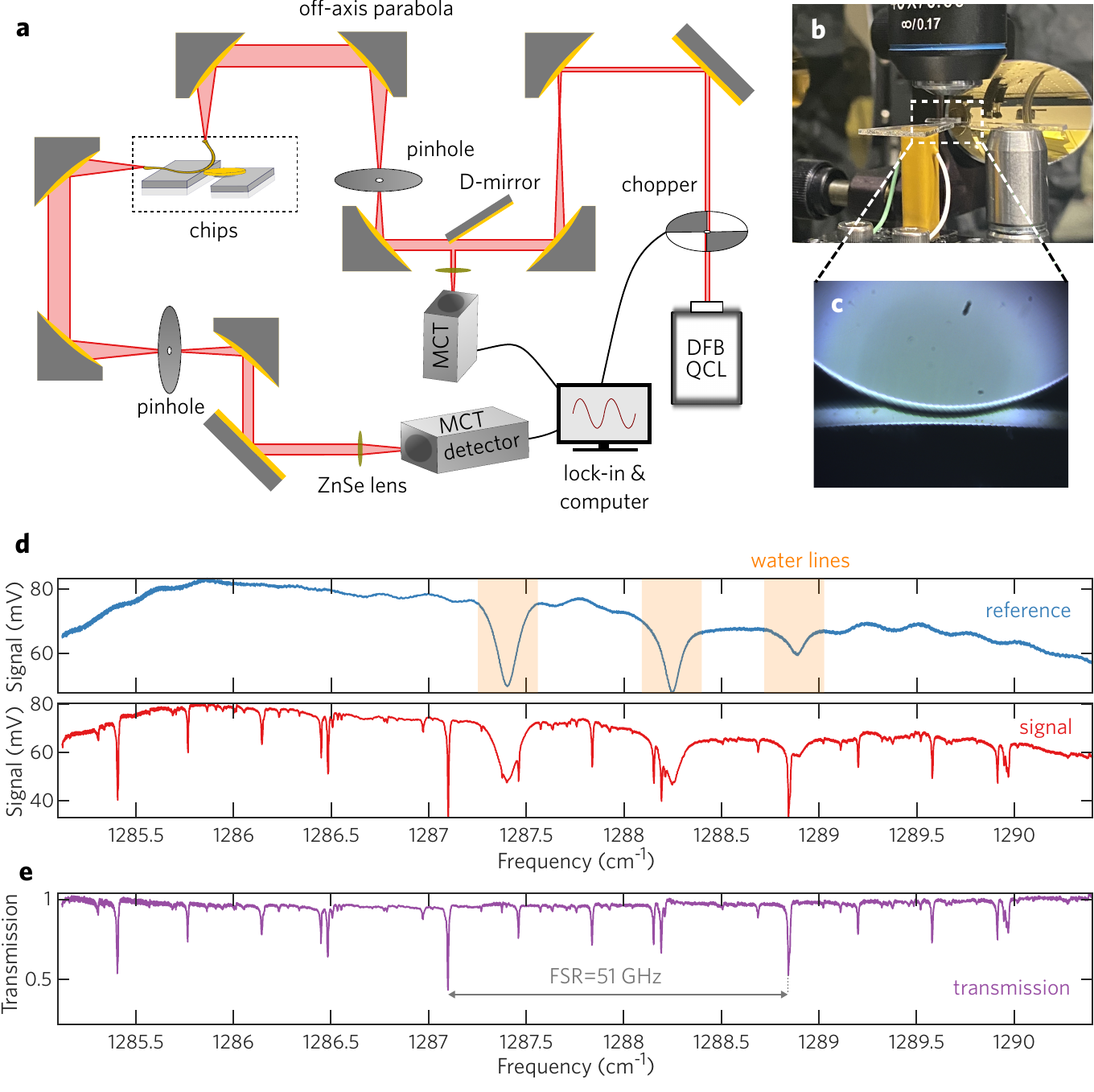}
    \caption{\label{fig:exp}\textbf{Experimental setup for transmission measurements.} \figsub{a} Schematic of the optical path and components. A DFB QCL is focused into the waveguide and coupled to the resonator. \figsub{b} Photograph of the coupling setup. \figsub{c} Optical image of the Ge WGM and waveguide when coupled, taken from the objective above the coupling area shown in \ref{fig:exp}b.}
\end{figure*}
To characterize the resonators, we performed direct transmission measurements using a single-mode distributed feedback QCL (DFB QCL).
% Previous work relied on Fouier transform spectroscopy and/or thermo-optically To characterize the \emph{Q} factor of the microresonators at LWIR, both Fourier transforms and thermal tuning has been applied to extract the change of waveguide transmission due to coupling with the microresonator\cite{kozak2021germanium,lee2020high}. In this study, we would like to establish a direct measurement using a single-mode distributed feedback (DFB) QCL laser.
The wavelength of the QCL can be tuned from 1285.5 to 1290.5 $\si{\cm^{-1}}$ by controlling both the QCL current and temperature. As is shown in the schematic of the optical path and components in Fig. \ref{fig:exp}a, the collimated laser beam was expanded through two off-axis parabola (OAP) pairs by a factor of ten and two times, consecutively, reaching a beam size of about 30 mm. Next, the collimated beam was focused on the fabricated Ge waveguide entrance facet using 1 inch OAP mirror, photographed in Fig. \ref{fig:exp}b, reaching a beam size of about $20\times20\,\, \si{\micro m^{2}}$. The transmitted laser signal from the waveguide was collected by an OAP of 4-inch focal length, and the collimated beam size was reduced by two times using a pair of OAPs. The transmitted signal was collected using ae Mercury-Cadmium-Telluride (MCT) detector connected to a lock-in amplifier system to eliminate background. We further used two pinholes at the confocal planes of the OAP mirror pairs to exclude unwanted interference from the scattered laser signal in the system. The waveguide chip was mounted on a piezoelectric actuator that allowed fine adjustment of the distance between the waveguide and the resonator.

Fig. \ref{fig:sim}d shows a pair of representative scans of waveguide with and without the coupling to the WGM microresonator. In the reference scan, the three dips of the transmission signals match well to the \ce{H_{2}O} gas lines at 1287.40 $\si{cm^{-1}}$, 1288.25 $\si{cm^{-1}}$, 1288.89 $\si{cm^{-1}}$\suppfig{S4}. We use these three gas absorption features as fine calibration for the QCL wavelength. When the WGM microresonator is coupled to the waveguide, we observed a series of sharp dips much narrower than the three \ce{H_{2}O} gas absorption lines. These come from coupling between the waveguide and the WGM microresonator. To acquire the transmission information without interference from the signal variation of the background, we divide the signal by the reference, and the normalized transmission is plotted in Fig. \ref{fig:exp}e. The most prominent dip corresponds to coupling to the fundamental whispering gallery mode, and it repeats three times when the laser is scanned from 1285 $\si{\cm^{-1}}$ and 1290 $\si{\cm^{-1}}$. This indicates a free spectral range of 51 $\si{\GHz}$. This value matches well with the calculation based on the FEM simulation, proving that the dips come from the coupling between the microresonator and the waveguide. Smaller dips are present as well (with the same FSR), indicating coupling to modes other than the fundamental. Coupling to these modes can be strengthened by adjustment of the waveguide's radial position. From the FSR, we estimate an effective group index of 4.15.

\begin{figure*}[h]
\centering\includegraphics{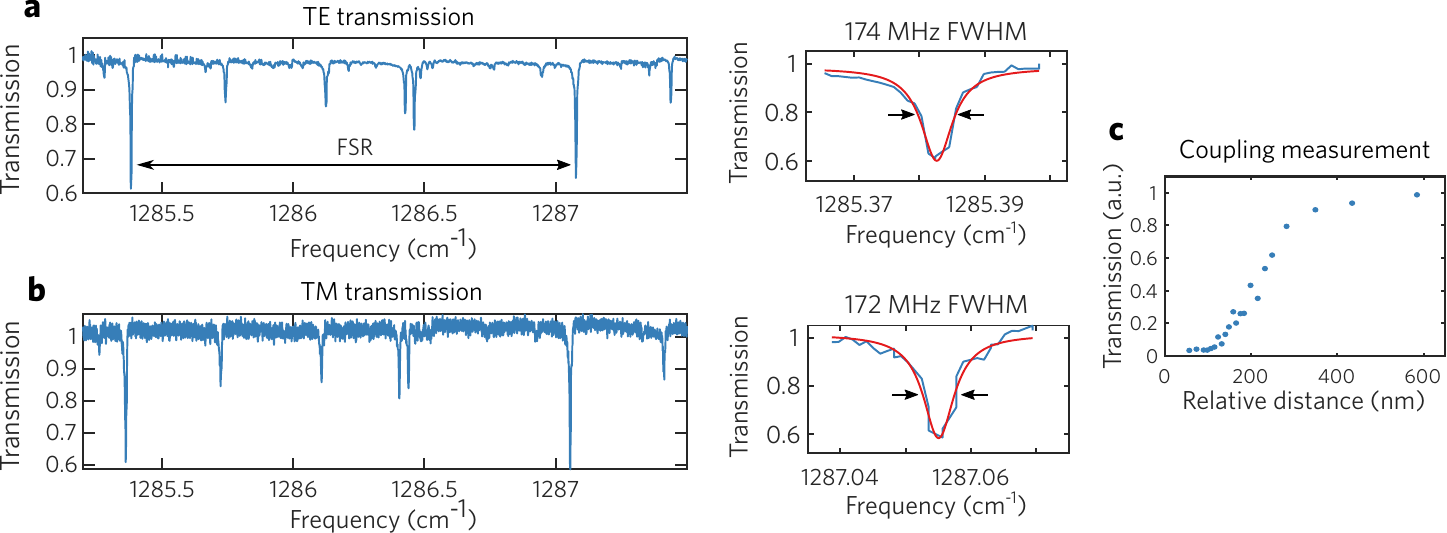}
\caption{\label{fig:resonators}\textbf{Polarization-dependent coupling measurements.} \figsub{a} TE polarization transmission, with a FWHM linewidth of 174 MHz.\figsub{a} TM polarization transmission, with a FWHM linewidth of 172 MHz. \figsub{c} Trend of transmission versus the change of vertical distance between the microresonator and waveguide. Due to the very high index mismatch between germanium and air, critical coupling is achieved only within a very short distance.}
\end{figure*}

To evaluate the properties of the WGM microresonator, the orientation of the QCL was adjusted to generate either TE- or TM-polarized laser emission, and this was fed into the optical path for the transmission measurement. Figs. \ref{fig:resonators}a and \ref{fig:resonators}b show the result of both the TE and TM transmission measurements, respectively, and at this distance each show two central transmission dips of about 60$\si{\%}$ separated by an FSR. A Lorentzian model was used to fit the dips at 1285.4 $\si{\cm^{-1}}$ and 1287.1 $\si{\cm^{-1}}$, giving a full width half maximum (FWHM) values of 174 GHz and 172 GHz, respectively. This corresponds to a loaded $Q_{l}$ of 221,000 and 224,000 for the respective TE and TM polarizations. To figure out the coupling condition, we performed a series of transmission measurements by changing the piezo stage's voltage, which in turn changes the separation between the WGM microresonator and the waveguide. Figure \ref{fig:resonators}c shows the transmission versus relative waveguide distance, and we can see a monotonic increase of the transmission over the range we can measure. This suggests that the 60\% transmission is in an under-coupled condition. That transmission dips corresponding to higher-order modes with \emph{FSR} smaller than 51 GHz start to grow significantly near the critical coupling point \suppfig{S5}, suggesting severe parasitic coupling between the waveguide mode and the higher-order WGM modes. The intrinsic \emph{Q} factor can be calculated using\cite{millerLowlossSiliconPlatform2017}
\begin{ceqn}
\begin{align*}\label{eqn:intrinsicQ}
Q_{i} = \frac{2Q_{l}}{1+\sqrt{T}}\tag{1}
%\label{eq1}
\end{align*}
\end{ceqn}
where $Q_{l}$ is the loaded \emph{Q} factor and T is the transmission when undercoupled. Using this, we calculate intrinsic \emph{Q} factors of 251,00 and 253,000 for the TE and TM polarization, respectively. There is little difference between these \emph{Q}s, suggesting that all surfaces of the WGM microresonators are equally smooth. These high-\emph{Q} values are almost two orders of magnitude higher than those from previous microresonators in the LWIR, and are summarized the fabrication method and its properties in Table \ref{tab:freq}. 

\begin{table}[!htb]
\captionsetup{size=footnotesize}
\caption{Comparison of previous LWIR microresonators.} \label{tab:freq}
\setlength\tabcolsep{0pt} % let LaTeX compute intercolumn whitespace
\footnotesize\centering
\smallskip 
\begin{tabular*}{\columnwidth}{@{\extracolsep{\fill}}cccc}
    \toprule
    Material & Synthesis & Type & Q factor
    \\ \midrule
    Diamond\cite{leeHighQDiamondMicroresonators2020} & CVD & WGM & 3,648\\
    SiGe\cite{ramirez2019broadband} & LEPECVD & Ring resonator & 3,200  \\
    Ge\cite{kozakGermaniumonsiliconWaveguidesLongwave2021} & MBE & Ring resonator & ~ 2,000 to 10,000 \\
    This work & Czochralski &  WGM & ~ 250,000
    \\ \bottomrule
\end{tabular*}
\end{table}
% Total circumference of the rings in Kozak et al: 1970.8 microns

From these intrinsic \emph{Q}s, we can now directly estimate the effective loss of the mode using $\alpha=\frac{2\pi n_g}{Q\lambda_0}$, giving an intrinsic loss of 0.13 \si{cm^{-1}} (0.5 dB/cm).
% 10 * log10(exp(((-(2 * pi * ((((51 GHz) / c) * ((2 * pi * (450 microns)) / 2))^(-1)))) / (253 000 / (1 288.5 (cm^(-1))))) * cm)) = -0.577851469
This value is very close to the reported losses of germanium in this range, showing that these resonators essentially have the highest qualities that are possible for the material (and possibly any material). The loss of the suspended portion of the waveguides is likely similar, although the larger losses of the glass-clad portion makes the averaged measured losses to be 9 \si{cm^{-1}}.

An attractive application for this platform is for chip-scale ultrabroadband sources pumped by QCLs, either by Kerr comb generation (in the resonator geometry) or by supercontinuum generation (in the waveguide geometry). While combs can be directly generated by QCLs,\cite{hugiMidinfraredFrequencyComb2012,luHighPowerFrequency2015,hillbrandCoherentInjectionLocking2019,mengMidinfraredFrequencyComb2020} achieving spectral bandwidths over 100 \si{cm^{-1}} is very challenging. The extremely wide transparency range of germanium makes it attractive for this application, as with it octave-spanning operation could be feasible. While in this work we did not attempt Kerr frequency comb generation---the inefficient coupling into the waveguide and the limited power of the QCL precluded this---it is worth calculating the parametric threshold for frequency comb generation. Assuming a modal diameter of 5 \micro m, we calculate that the threshold for comb generation should be approximately 240 mW under optimal conditions\cite{xuanHighQSiliconNitride2016}. 
% ((((1 + .5)^3) / .5) * (pi^2) * (4^2) * ((450 microns) / 2) * pi * ((2.5 microns)^2)) / (((2 * (20 (nm^2))) / W) * (1 / (1285 (cm^(-1)))) * (250 000^2)) = 0.242046623 watts
This threshold is well below the 10 W range produced by state-of-the-art LWIR QCLs \cite{zhouHighpowerContinuouswavePhaselocked2019}. Still, achieving fully-integrated comb generation in the current implementation of this platform would be difficult due to the lift-off required. Though the waveguides are fairly lossy when on the glass, they are actually remarkably low when one considers that the losses of glass can exceed 10,000 \si{cm^{-1}} in the LWIR. This is a result of the large index mismatch of germanium and the glass, leading to a small overlap of the mode with the glass. Predeposition of a thin film of a moderately low-loss low-index material, such as ZnS, would drastically reduce the loss to a similar level as the disks. 

In conclusion, we presented a non-epitaxial fabrication platform for germanium microresonators and waveguides whose quality factors are two orders of magnitude higher than prior work. This methodology avoids the defects inevitably associated with thick epitaxial growth. Using a single-mode DFB QCL, we observed transmission dips with repetition rate matched well to the FSR of the WGM microresonator. Loaded \emph{Q} factors of 221,000 and 224,000 were measured for the TE and TM polarization, respectively, corresponding to the intrinsic \emph{Q} factors of $2.5 \times 10^{5}$. These excellent \emph{Q} values represent a milestone for the low-loss LWIR integrated photonics, and we expect similar platforms to be essential for future chip-scale sensing and nonlinear photonics.    

\ifpreprint \else
    \section*{Methods}
    \subsection*{Wafer bonding process} \waferbonding
    \subsection*{Nanofabrication process} \nanofab
    \subsection*{Tuning of the quantum cascade laser} \dfbtuning
    \subsection*{Contributions}
    D.R. and D.B. initiated the project. D.R. fabricated the devices, performed the simulation and measurements. D.R. and D.B. designed the measurement setups.  D.R. and C.D. performed the experiments. D.R. wrote the manuscript with contribution from all authors. D.B. supervised the whole project.
\fi

\subsection*{Acknowledgements}
This work is partly sponsored by Research Council of Norway through the FRINATEK Program (Grant No. 302923). D.R. thanks the generous financial support from the FRIPRO Mobility Fellowship, Research Council of Norway. D.B. acknowledges support from AFOSR grant no. FA9550-20-1-0192, NSF grant ECCS-2046772, and ONR grant N00014-21-1-2735.

\bibliography{ref.bib}
\end{document}